\documentclass[twocolumn,pra, aps,superscriptaddress,showpacs,groupedaddress]{revtex4-1}

\usepackage{mathptmx}
\usepackage{subfigure}
\usepackage{dcolumn}
\usepackage{amsmath,amssymb}
\usepackage{bm}
\usepackage{color}
\usepackage{latexsym}
\usepackage{epstopdf}
\usepackage{color}
\usepackage[english]{babel}
\usepackage{latexsym}

\usepackage{psfrag,graphicx} 
\usepackage{epsf} 
\usepackage{subfigure} 
\usepackage{amsmath} 
\usepackage{amssymb} 
\usepackage{amsfonts}
\usepackage{bm}
\usepackage{natbib}
\usepackage{epstopdf}
\usepackage{appendix}

\definecolor{mygrey}{gray}{0.35}
\definecolor{myblue}{rgb}{0.2,0.2,0.8}
\definecolor{myzard}{cmyk}{0,0,0.05,0}
\definecolor{mywhite}{rgb}{1,1,1}
\definecolor{myred}{rgb}{1,0.,0.3}

\usepackage[colorlinks=true,citecolor=myblue,linkcolor=myred]{hyperref}

\def\be{\begin{equation}}
\def\ee{\end{equation}}
\def\ba{\begin{align}}
\def\enda{\end{align}}
\def\bi{\begin{itemize}}
\def\ei{\end{itemize}}

\def\N{{\mathbb{N}}}

 \def\ee{\mathord{\rm e}}

\def\N{{\mathbb{N}}}

 \def\ee{\mathord{\rm e}}

\newcommand{\Yb}{$^{171}{\rm{Yb}}^{+} $}

\renewcommand{\ee}{{\rm e}}

\def\beq{\begin{equation}}
\def\beq{\begin{equation}}
\def\eeq{\end{equation}}

\newcommand{\bla}[1]{\left(#1\right)}
\newcommand{\blb}[1]{\left[#1\right]}

\begin{document}

\title[Short Title]{Refocusing two qubit gate noise for trapped ions by composite pulses}

\author{I. Cohen}
\thanks{These two authors contributed equally}
\author{A. Rotem}
\thanks{These two authors contributed equally}
\author{A. Retzker}
\affiliation{Racah Institute of Physics, The Hebrew University of Jerusalem, Jerusalem 91904, Givat Ram, Israel}


\begin{abstract}
{Amplitude noise which inflicts a random two qubit term is one of the main obstacles preventing the implementation of a  high fidelity two-body gate below the fault tolerance threshold. This noise is difficult to refocus as any refocusing technique could only tackle noise with frequency below the operation rate. Since the two qubit gate speed is normally the slowest rate in the system, it constitutes the last bottleneck towards an implementation of a gate below the fault tolerant threshold.  Here we propose to use composite pulses as a dynamical decoupling approach, in order to reduce two qubit gate noise for trapped ions systems. This is done by refocusing the building blocks of ultrafast entangling gates, where the amplitude noise is reduced to shot-to-shot (STS) noise. 
We present detailed simulations showing that the fault-tolerance threshold could be achieved with the proposed approach.
}
\end{abstract}

\maketitle

\newpage

\section{Introduction}
Fault-tolerant quantum computation\cite{knill, aharonov,kitaev,Aliferis}  has been considered to be a key goal of the field of quantum information science. 
In order to realize quantum gates with arbitrary precision 
single-body and two-body operations should be performed below the fault-tolerance threshold, where noise and other decoherence effects must be compensated. To this end, methods such as dynamical decoupling have been utilized in different quantum architectures, e.g. trapped ions, super-conducting qubits and NV centers, performing single-body operations below the threshold \cite{Harty2014prl,Barends2014Nature,Rong2015Naturec}. However, fault-tolerant two-body operations are yet to be demonstrated. 

In order to realize fault-tolerant two-body operations, a few obstacles should be overcome. Two dominant ones are relaxation originating from a decaying excited state, and amplitude noise of the interaction due to Rabi frequency fluctuations. In trapped ion systems, relaxation originating from a decaying dipole transition is reduced drastically using Raman transitions, providing the ability to reduce photon scattering by using high power lasers \cite{Roee1}.
For this reason high power pulsed lasers are considered extremely promising \cite{Mizrahi2013prl,Mizrahi2}.
 However, it still remains unknown how to reduce the amplitude noise of two qubit gates. 
In this paper we address this obstacle, namely we show that we can counter the amplitude noise in the two-body interaction using the method of composite pulses, especially for pulsed lasers. By this we substantially facilitate the route towards fault tolerance.

In general, amplitude noise could be tackled using composite pulse sequences. Composite pulses have been first invented for nuclear magnetic resonance implementations \cite{NMR1,NMR2,NMR3}, to counter small systematic errors $\epsilon$ in the pulses. This approach has been considered in the context of quantum computation \cite{Brown2004pra,Morton2005prl}, where using fully compensating composite pulses the systematic error is reduced to an arbitrary order $O(\epsilon^n)$, regardless of the qubit's initial state. 
The composite pulses method for compensating amplitude errors in single-body rotations can be extrapolated to correct any set of operations that share the same algebra of Pauli matrices, 
e.g., two-body interactions  \cite{Jones2003pra1,Hill2007prl,Testolin2007pra,Tomita2010njp,Ichikawa2013pra}, where the rotational SU(2) group can be mapped to a subgroup of SU(4). 

Recently, composite pulses have been considered to compensate time dependent $1/f$ noise of single-body operations \cite{Wang2014pra,Kabytayev2014pra}, where the composite pulses can be thought of as dynamical decoupling. We stress that applying this approach directly on the two-body interaction term would not work in state of the art experiments with trapped ions due to the fact that the gate speed is lower than the typical noise frequency. Thus, the only way to tackle this issue is to correct the fundamental gate ingredients which are much faster. To this end,  we show that we can refocus the elementary building block of the ultrafast gate, resulting in a refocused two-body interaction. 

\section{Single-body refocusing approach}
The composite pulses method has been initially considered for refocusing a single-body operator with a systematic error \cite{NMR1,NMR2,NMR3,Brown2004pra,Morton2005prl}.  For example,
\beq H=g\bla{1+\epsilon}\sigma_x, \label{H0}\eeq 
where $\epsilon$ designates the noise, $g$ is the gate amplitude and $\sigma_x$ denoting the Pauli matrix in the $x$ direction. The aim is to perform the operation with arbitrary accuracy, namely to realize the Hamiltonian $H=g\sigma_x+O(\epsilon^n).$
For simplicity, the case where the first order of the error is countered, is presented here.  
The basic idea is to use the same Hamiltonian in eq.\ref{H0} for inducing two additional rotated unitaries: $U_{\alpha_1}(\tau_1)=\exp\bla{ig[1+\epsilon]\sigma_{\alpha_1} \tau_1}$, and $U_{\alpha_2}(\tau_2)=\exp\bla{ig[1+\epsilon]\sigma_{\alpha_2} \tau_2}$, for time durations $\tau_1=\tau_2=2\pi/g$ and Pauli matrices $\sigma_{\alpha_1}=\cos\theta \sigma_x  +\sin\theta \sigma_y$ and  $\sigma_{\alpha_2}=\cos\theta \sigma_x  -\sin\theta \sigma_y$. 
Thanks to the periodicity, the product of these two unitaries results in 
\beq
U_2(\tau_2) U_1(\tau_1)=\exp \bla{i2g\cos\theta\epsilon\sigma_x}+O(\epsilon^2).
\eeq
Therefore, concatenating all the three unitary $U_2(\tau_2) U_1(\tau_1) U(\tau)$ results in countering the first order of the error, after setting $2\pi\cos\theta=g\tau$. By using a more elaborate scheme, it is also possible to eliminate higher orders in the error.

Now, instead of having a systematic error $\epsilon$ we could rather think of having a noisy term with $\epsilon(t)$, the above scheme can be thought of as dynamical decoupling, which is performed appropriately as long as the whole operation takes less time than the noise's correlation time, i.e. under the assumption that the noise is not changing during the whole operation $\epsilon(t)=const$. In what follows, we show an extrapolation of the single-body refocusing approach, for the refocusing of a noisy interaction.



\section{Refocusing interaction $H=g\sigma_z^1\sigma_z^2$}
In many quantum systems one of the leading sources of noise is the fluctuating Rabi frequency of the driving fields. This yields a single-body noisy term, which can be dealt using the approach taken above. Indeed, it was shown experimentally that single-body operation can reach the fault tolerance threshold \cite{Harty2014prl}. However, Rabi frequency fluctuations are also responsible to an amplitude noise of the two-body interaction 
 \beq
H=g\bla{1+\epsilon(t)}\sigma_x^1\sigma_x^2.
\label{inter}
\eeq
In order to refocus the the noise in the interaction \cite{Jones2003pra1,Hill2007prl,Testolin2007pra,Tomita2010njp,Ichikawa2013pra}, one can use an extrapolative version of the single-body refocusing approach, which only requires the periodicity of the time propagation unitary of the Hamiltonian, namely $U(t_k)=\exp\bla{-ig\sigma_x^1\sigma_x^2 t_k}=1$, for times $t_k=2\pi k/g$ and integer $k\in\N$. Similarly to the single-body case, when operating with the Hamiltonian in eq.\ref{inter} for  time $\tau$, the desired unitary is $U(\tau)=\exp\bla{-ig\sigma_x^1\sigma_x^2 \tau}$, whereas the undesired one $U^{UD}(\tau)=\exp\bla{-ig\epsilon(\tau)\sigma_x^1\sigma_x^2 \tau}$ should be countered. As before, this can be done by concatenating two additional rotated unitary, originating from eq. \ref{inter}: $U_1=\exp\bla{ig[1+\epsilon(\tau)]\sigma_x^1\sigma^2_{\alpha_1} \tau_1}$, and $U_2=\exp\bla{ig[1+\epsilon(\tau)]\sigma_x^1\sigma^2_{\alpha_2} \tau_2}$, for time durations $\tau_1=\tau_2=2\pi/g$ and Pauli matrices $\sigma_{\alpha_1}=\cos\theta \sigma_x  +\sin\theta \sigma_y$ and  $\sigma_{\alpha_2}=\cos\theta \sigma_x  -\sin\theta \sigma_y$. Note that rotating the Pauli matrices in these unitary with respect to eq. \ref{inter}, is achieved by single-body operations, which are assumed to have high fidelities. Using the periodicity of these unitary, their product results in 
\beq
U_2 U_1=\exp \bla{i2\cos\theta\epsilon(\tau)\sigma_x^1\sigma^2_x}+O(\epsilon(\tau)^2).
\eeq
Therefore, concatenating $U(\tau)$ with these two rotated unitary, results in countering the noise to the first order, by setting $2\pi\cos\theta=g\tau$. Importantly, the assumption that the noise $\epsilon(t)$ does not change during the whole operation must hold. Meaning, that the power spectrum of the noise at the pulse rate frequency still inflicts loss of fidelity. However, two-body interactions are usually much weaker than the single-body operations, and unfortunately, might last longer than the time correlation of the noise. Therefore, the inevitable assumption might not hold in all cases, especially when higher orders of the noise are compensated by even longer dynamical decoupling operations.


{\it Refocusing of the building blocks of ultra-fast entangling gates.--- }
In order to appropriately use the dynamical decoupling approach for compensating the noise in the interaction only terms which are faster than one over the noise correlation time  could be refocused.  Thus,  instead of refocusing the two qubit gate it is possible to refocus the main building blocks of the gate, which are much faster, and then construct the gate. For trapped ions, several schemes of implementing ultra-fast entangling gates have been proposed, and thus can be used in the current proposal for the refocusing of their amplitude noise. However, since these schemes make the use of pulsed lasers, the main noise is a shot-to-shot (STS) noise, i.e., each pulse makes a rotation at a different phase with no correlation between the pulses. Therefore, it would not suffice to counter the amplitude noise of the whole gate, as was demonstrated above, in the case of the slow Molmer Sorensen gate\cite{MS}. Instead, we should use the dynamical decoupling approach for refocusing the ultra-fast gate's building blocks, which are generated from the same pulse of the laser, namely, having the same shot noise.

In ref. \cite{Garcia2004}, Garcia-Ripoll {\it et al.} have proposed to use a single building block for implementing the ultrafast entangling gate, which is realized by the following Hamiltonian: 
\beq
\begin{split}
H_{BB} = \Omega \sum_{i=1,2}\sigma^i_+e^{ i\eta\bla{b^\dagger+ b}} + h.c 
\label{bb1}
\end{split}
\eeq 
where $\eta$ is the Lamb-Dicke parameter, $b^\dagger$ ($b$) are the creation (annihilation) of a vibrational phonon, and we have neglected the vibration term, since the motional dynamics during the pulse is negligible. For a very short pulse duration $\tau=\pi/\Omega$ the achieved building block unitary is the following spin dependent kick (SDK)
\beq
\begin{split}
U_{BB} =\exp\blb{-i\frac{\pi}{2}\sum_{i=1,2}\sigma^i_+e^{ i\eta\bla{b^\dagger+ b}} + h.c}.
\label{bb2}
\end{split}
\eeq 
It turns out that the building block fulfills the conditions needed for refocusing using the extension of Brown's {\it et al.},\cite{Brown2004pra} approach if the impulsive limit is assumed. Meaning, during the pulse sequence the  $b,b^{\dagger}$ operators are frozen and could be looked upon as scalars. Thus the operator in eq.\ref{bb1} is just the Pauli operator and we are back to the same scenario as shown after eq.\ref{H0}.  Alternatively, it could be seen that the following criteria are valid: (i) The Hamiltonian in eq.\ref{bb1} can be rotated by a single-body $\sigma_z$ operation. (ii) As the operator in eq.\ref{bb1} squares to identity, the unitary in eq.\ref{bb2} is periodic, i.e., by applying an even number of such pulses, we obtain the unity $\mathbb{I}$.

When introducing the STS noise, we assume that a sequence of five pulses,  originating from a single laser pulse, have the same Rabi frequency noise, such that the unitary of each of the five pulses is
\beq
\begin{split}
U^j_{BB} =\exp\blb{-i\bla{\frac{\pi}{2}+\epsilon}\sum_{i=1,2}\sigma_+^i e^{i\theta_j}e^{ i\eta\bla{b^\dagger+ b}} + h.c} 
\label{im}
\end{split}
\eeq 
where $\epsilon$ is the equivalent shot noise of all the five pulses. By generating a single-body $\sigma_z$ rotations, each $j^{th}$ pulse accumulates additional $\theta_j$ phase, such that $\theta_0=0$, $\theta_1=\theta_2=\theta$, and $\theta_3=\theta_4=-\theta$. Using the periodic condition, the concatenations of pulses $2$ with $3$ and $4$ with $5$ result in 
\beq
\begin{split}
U^{2}_{BB}U^{1}_{BB} =-\exp\blb{-i2\epsilon\sum_{i=1,2}\sigma_+^i e^{i\theta}e^{ i\eta\bla{b^\dagger+ b}} + h.c}\\
U^{4}_{BB}U^{3}_{BB} =-\exp\blb{-i2\epsilon\sum_{i=1,2}\sigma_+^i e^{-i\theta}e^{ i\eta\bla{b^\dagger+ b}} + h.c}\\
\end{split}
\eeq 
The concatenation of all four rotated unitary yields $\exp\blb{-i4\epsilon\cos{\theta}\sum_{i=1,2}\sigma_+^i e^{ i\eta\bla{b^\dagger+ b}} + h.c} + O(\epsilon^2)$, thus compensating the first order of the STS noise, by setting $\cos\theta=-1/4$. 

\section{Ultrafast gate for Hyperfine qubits}
In the derivation of eq. \ref{bb1}, the rotating wave approximation (RWA) was taken, namely the Rabi frequency $\Omega$ is assumed to be much smaller than the quadrupole energy gap $\omega_{qu}$, such that, $\omega_{qu}\tau \gg 1$ must hold. Exchanging the quadrupole ions with hyperfine ones, whose energy gap is orders of magnitude smaller than the quadrupole energy gap, imposes the use of longer pulse durations. In that case, the vibrational dynamics (VD) might not be negligible during the longer operation duration $T=\tau$, thus resulting in a reduced fidelity which scales as $IF\propto (\nu T)^2$. This effect is more dominant when applying the five pulse sequence for the STS refocusing, where $T\approx 5\tau$. In the following we give a careful treatment for the refocusing of building-block fidelity. 

We consider trapping the \Yb ions with secular frequencies in the proximity of $\nu/2\pi \approx 1$ MHz. The \Yb ion has hyperfine energy gap of $\omega_{hf}/2\pi=12.6$ GHz, thus a pulse duration $\tau > 10$ $ns$ 
should be applied in order to justify the RWA taken in eq. \ref{bb1}. During this relatively long pulse duration, the VD is not negligible anymore, thus in the rotating frame of the vibration, $b^\dagger$ and $b$ accumulate a non-vanishing phase $\pm \nu\tau/2\pi  \approx \pm10^{-2}$ respectively, resulting in continuously changing the SDK's direction in phase space, 
\beq
\begin{split}
U_{BB} =T \exp\blb{-i\int_0^t dt' \sum_{i=1,2}\frac{\Omega(t')}{2}\sigma^i_+e^{ i\eta\bla{b^\dagger e^{i\nu t'}+ h.c}} + h.c},
\label{bb2rotated}
\end{split}
\eeq 
with $T$ being the time ordering, and the pulse's Rabi frequency is $\Omega(t)=\frac{\pi}{\tau} \mbox{sech} (\pi t/\tau),$ \cite{Mizrahi2}. In the above eq. the time dependent displacements do not commute in different times, thus dropping the fidelity of achieving the SDK of eq. \ref{bb2}. 

\begin{figure}[h]
\begin{center}
\hspace{-0.25cm}
\includegraphics[width=0.45\textwidth]{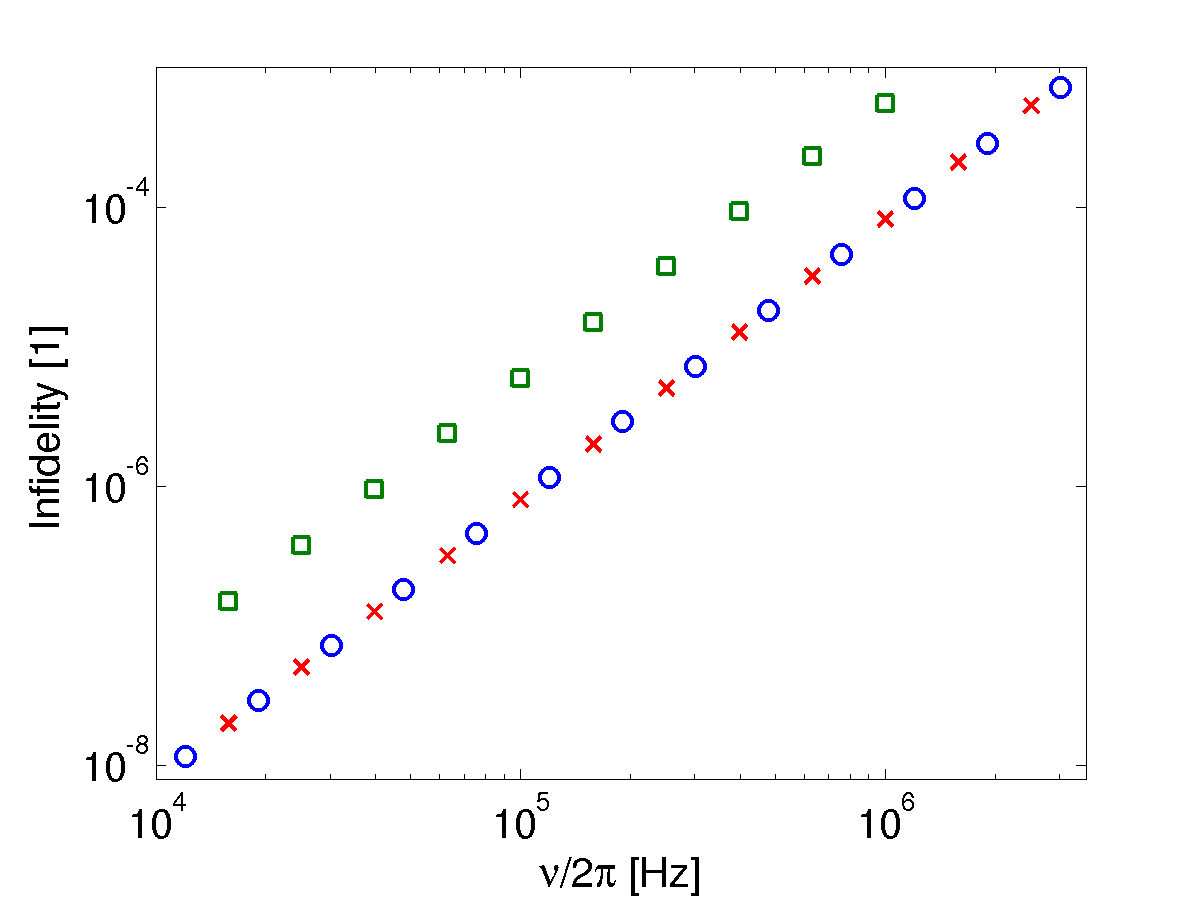}
\end{center}
\caption{
{\bf Infidelity of a SDK as a function of the trap frequency, neglecting the STS noise.} In the first approach for generating the ultrafast SDK, if we neglect the STS noise, the single source of infidelity is the VD. Thus, by applying a single pulse (blue curve), the infidelity scales as $\propto \bla{T\nu}^2$, where $T$ is the operation duration. By composing five SDKs (green curve), the time duration of the whole refocused SDK operation is increased, therefore, the fidelity is reduced. Taking the VD effect into account, by adjusting the phase and the magnitude of the displacement of the whole five composite SDKs together, (red curve) the fidelity is increased, reaching the fidelity value of a single pulse, thus, compensating the undesired effect. 
}
\label{trap_dependence_A}
\end{figure}

\begin{figure}
\begin{center}
\hspace{-0.25cm}{A}
\includegraphics[width=0.4\textwidth]{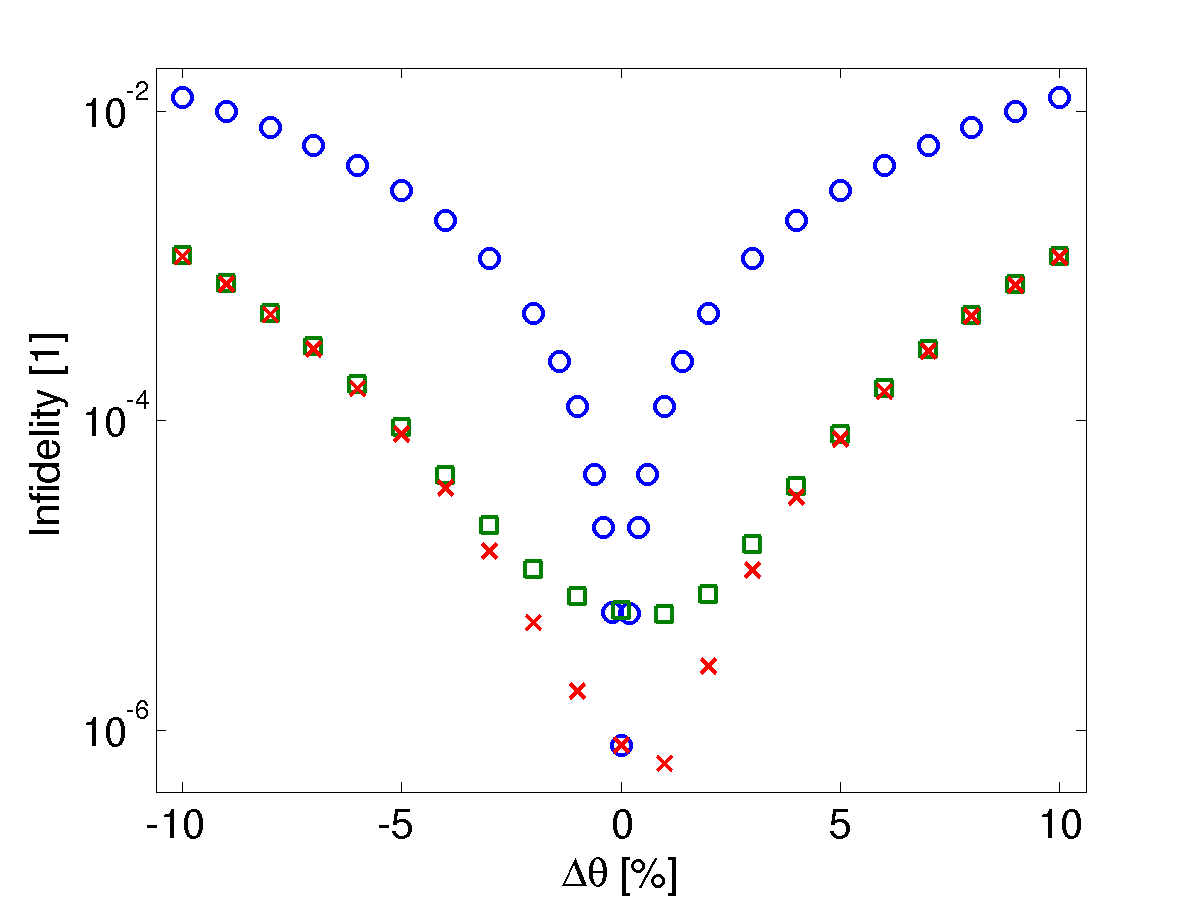}
\hspace{-0.25cm}
{B}\includegraphics[width=0.4\textwidth]{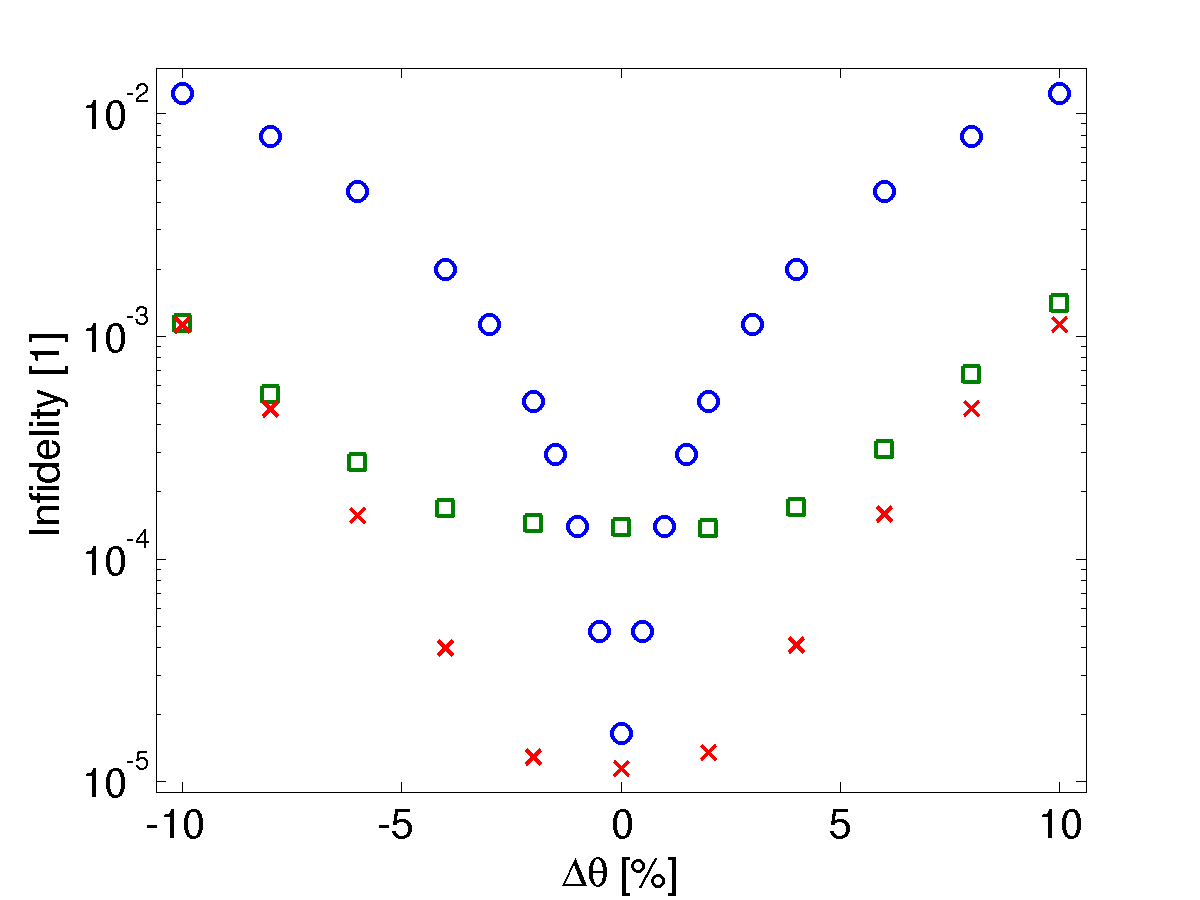}

\end{center}
\caption{{\bf Infidelity due to STS noise simulated for $\nu/2\pi=100$ kHz.} In both approaches for generating a SDK by applying a single pulse (A blue) or a single pulse train (B blue), the infidelity raises rapidly with the STS noise. Using the composite pulse approach to refocus the SDK, by applying five SDK pulses (A green) or five pulse trains (B green), the STS noise is reduced; yet, due to the non negligible VD the fidelity is lower than the single SDK (blue). Partially compensating for the undesired VD effect is achieved by adjusting the phase and magnitude of the SDK's displacement. This improves the fidelity of the five composite pulses (A red) or trains (B red), while still being decoupled from the STS noise. A better behavior is seen in the first approach due to the additional errors of the constructive interference of the second approach.  
}
 \label{shot1}
\end{figure}

Using numerical simulation of the dynamics, we evaluate the SDK fidelity by $F=Tr\bla{U_{eff}^\dagger U_{sim}  }$, where $U_{eff}$ and $U_{sim}$ are the desired and simulated unitary respectively. The cutoff of the vibrational phonon is 30, and for the fidelity evaluation we consider only the first 7 Fock phononic states \cite{assumption}. According to our numerical simulations, as long as $\nu\tau/2\pi < 10^{-2}$, the SDK in eq. \ref{bb2rotated} has an infidelity of $IF=10^{-6}$ (fig. \ref{trap_dependence_A}), since the motional dynamics is negligible. Therefore, by concatenating 28 such pulses, having alternating displacements, the two-qubit phase gate generated \cite{Mizrahi2,Garcia2004} yields infidelity $IF<10^{-4}$ below the threshold. However, in this naive treatment, we have not considered the STS noise. Including this effect, the fidelity of a single SDK drops (fig. \ref{shot1}A). Compensating the STS noise by using the five composite SDKs now takes a longer time, thus due to the non negligible VD a reduction in the fidelity occurs (fig. \ref{trap_dependence_A}). In order to increase the fidelity, we adjust the displacement's direction (phase) and strength, of the whole five composite pulses together. In the improved case, the achieved infidelity of the refocused SDK is $IF=10^{-6}$ for STS noise less than $3\%$ and for $\nu/2\pi<100$ kHz.

\begin{figure}[h]
\begin{center}
\hspace{-0.25cm}
\includegraphics[width=0.45\textwidth]{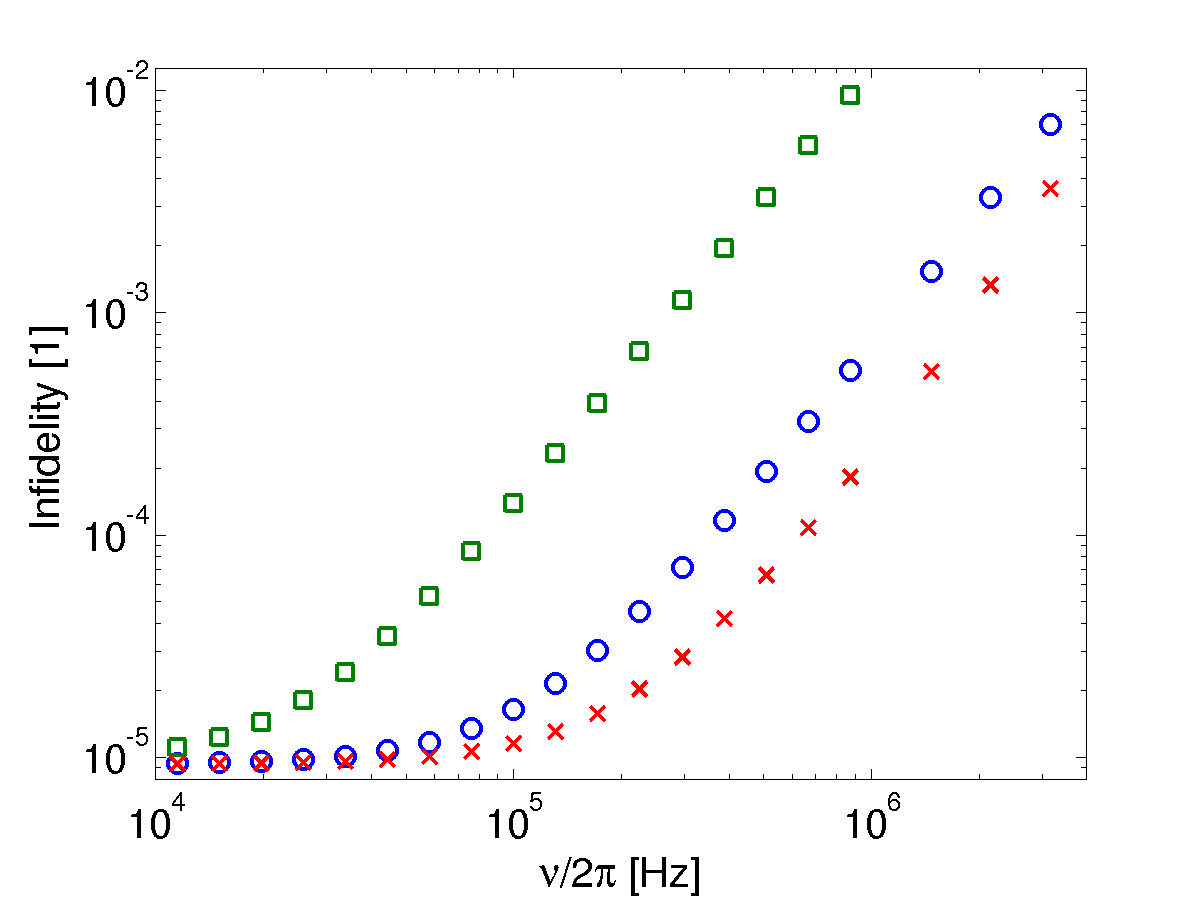}
\end{center}
\caption{
{\bf Infidelity of a SDK as a function of the trap frequency, neglecting the STS noise.} Generating the ultrafast SDK in the second approach, is done by applying a pulse train of $2^N$ counter-propagating pulse twins, with $N=8$. By applying a single SDK train,  (blue curve) the infidelity saturates at $O(\theta^2)\approx O(2^{-2N})\approx 10^{-5}$. Further away from the saturation regime, the infidelity scales as $\propto \bla{T\nu}^2$, similarly to the behavior of the first approach. Applying five composite trains to generate a refocused SDK yields a decreased fidelity due to the more dominant VD. By adjusting the displacement of the whole five composite SDK trains, (red curve), the VD effect is partially compensated, resulting in an increased fidelity, higher than the unadjusted single train case. 
}
\label{trap_dependence_B}
\end{figure}


\section{Different approach for generating ultrafast SDKs} 
In ref. \cite{Mizrahi2013prl} Mizrahi {\it et al.} have experimentally demonstrated a different way to generate an ultrafast SDK, using frequency-comb concept. In their experiment, a single pulse arriving from a pulse laser, is distributed into a train of $2^N$ equal pulses using a set of N beamsplitters imposing N delay lines. Each of these pulses is split once again to two twin pulses in separated arms, which are then focused on the trapped \Yb ions simultaneously, performing a Lin $\perp$ Lin polarized counter-propagating pulses. This configuration induces a polarization gradient \cite{lin_lin} to couple the qubit states with the vibration, such that every twin pulse couple yields 
\beq
\begin{split}
H =  \nu b^\dagger b + \frac{\omega_{hf}}{2}\sum_{i=1,2}\sigma_z^i+\sum_{i=1,2} {\Omega(t)}\sigma^i_x \sin \bla{ \eta\bla{b^\dagger + b}+\omega_A t},
\label{H_Monroe}
\end{split}
\eeq 
after adding to one of the Raman laser arms an acousto optic modulator with $\omega_A$ being its modulation frequency, and where the Rabi frequency is $\Omega(t)=\frac{\theta}{ \tau} \mbox{sech} (\pi t/\tau)$, and $\theta = \pi \mbox{sech}(\omega_{hf} \tau/2)/2^N$, with pulse duration $\tau=10$ ps, satisfying $\tau \omega_{hf} \ll 1$.

As apposed to the previous approach, this Hamiltonian is derived in the lab frame, without moving to the rotating frame of the hyperfine energy. Therefore, during the pulse duration the system evolves not only by the vibrational term and the pulse driving (first and last term of eq. \ref{H_Monroe} respectively), as is considered at the quadrupole ion system (eq. \ref{bb1}), but also according to the ion's energy gap (second term of eq. \ref{H_Monroe}). To preserve energy, that is to compensate the large energy mismatch, a constructive interference of the $2^N$ couples of twin pulses is built, such that the desired SDK (eq. \ref{bb2}) is generated, in the large N limit, where the infidelity saturates at $O(\theta^2)\approx O(2^{-2N})$, (fig. \ref{trap_dependence_B}). However, increasing the number of delay lines, also increased the time operation $T_{tot}\propto2^N$, thus due to the non negligible VD, the infidelity increases as $(\nu T_{tot})^2$. Operating with $N=8$ delay lines and trap frequency $\nu/2\pi < 100$ kHz yields infidelity of a single SDK below $IF<10^{-5}$, such that by concatenating 28 such SDK trains, with alternating displacements, an entangling gate with infidelity $IF<3\cdot 10^{-4}$ is generated \cite{Mizrahi2,Garcia2004}. Further improvement of the achieved gate fidelity may be enhanced by increasing $N$ the number of beam-splitters. In addition, when introducing the STS noise, the fidelity drops even more. To tackle this effect we can use the composite pulse scheme, which however, extends the operation duration, and therefore, the VD problem becomes more dominant. As was done above, by adjusting the refocused SDK's displacement, the VD effect is partially suppressed, while the STS noise is compensated (fig. \ref{shot1} B).



\section{Summary}
In this paper we have tackled the issue that refocusing techniques cannot be applied directly on two qubit gates as these are slower than the correlation time of the noise. We have concentrated on amplitude noise as this is one of the main bottlenecks of state of the art experiments.  As the composite pulses technique cannot be applied directly on the two qubit gate, we have applied it to refocus the building blocks of the ultrafast entangling gates, i.e., the SDKs, where the amplitude noise is reduced to the STS noise. We have shown that in both SDK generating approaches, an entangling gate can be generated with infidelity below the threshold. 
We believe that coherent control methods could improve further the presented results by efficient pulse shaping.


\section{ Acknowledgments} We would like to acknowledge the support of the ARO Quantum Computing Program.









\begin{references}
\bibitem{assumption} To verify the approximated SDK fidelity, we have simulated a complete ultrafast gate using $8$ SDKs, to close the phase space of a single vibrational mode. Initializing the system the vibrational mode ground state, the total fidelity behaves as expected $IF_{gate}\approx2\cdot 8 IF_{SDK}$, (2 for the two ions), verifying our assumption.
\end{references}
\end{document}